# Superfluids and superconductors - an 80 year perspective

W.P. Halperin

*Superfluids and superconductors have a common conceptual basis in systems ranging from the lightest to the heaviest.*

The origins of quantum condensed phenomena are traceable to the serendipitous discovery of superconductivity and superfluids in the early part of the last century where electrons and atoms were found to flow without any resistance over great distances. Such non-intuitive behavior is beautifully demonstrated by superfluid helium flowing through extremely narrow channels, reported in two landmark results published in 1938 in the journal Nature by Allen and Misener[1] and by Kapitza[2], the latter receiving the 1978 Nobel prize. Although the relationship to superconductivity as an inviscid phenomenon was recognized, neither had the firm theoretical understanding that exists today which provides the foundation to investigate new superfluid phases and topological superconductors.

Helium becomes a superfluid at the temperature 2.17 K, close to the condensation temperature predicted by Bose and Einstein for non-interacting particles. Shortly after the observations,[1,2] Fritz London offered a possible explanation[3] in terms of Bose-Einstein condensation. With foresight Kapitza also noted a possible connection to 'supraconductivity' for which a complete theory was eventually realized in 1957 by Bardeen, Cooper, and Schrieffer[4] (BCS) and were given the Nobel prize in 1972.

Fundamental to quantum particles of any type is their indistinguishability, and there are exactly two classes. They are either even or odd under particle exchange and are called bosons and fermions respectively. However, an even combination of interacting fermions can make a composite boson, as is the case for the six fermions that constitute the $^4$He atom (two electrons, two neutrons, and two protons). At sufficiently low temperatures the $^4$He boson quantum condenses into a superfluid. Similarly, following BCS, electrons in a metal with a suitably attractive interaction can form Cooper pairs. Although not a localized composite boson of electrons, this pairing is a condensation in their momentum space distribution resulting in the superfluid known as a superconductor.

In the wake of the second world war production of tritium for the hydrogen bomb made available significant quantities of the light helium isotope, $^3$He. Since $^3$He is a composite particle of an odd number of fermions it is not a boson and Bose-Einstein condensation cannot take place. One might guess that it could never be a superfluid. However, success of the BCS theory for superconductivity suggests another possibility. Perhaps a composite boson of Cooper pairs of $^3$He atoms might condense to a superfluid, much like the electrons of the BCS superconductor. This hypothetical superfluid was calculated by Anderson and Morel[5], later modified by Balian and Werthamer[6] and independently by Vdovin.[7] Then the subject exploded following an unexpected discovery of this novel superfluid at 0.0025 K by Osheroff, Richardson, and Lee in 1972 for which they received the Nobel prize in 1996.[8] At first their results were interpreted as spontaneous nuclear magnetic ordering in solid $^3$He, but correctly identified shortly thereafter as the transition to a superfluid.[9] In fact, nuclear magnetic ordering in the solid phase was discovered two years later, about three times lower in temperature with experiments extending to 0.0007 K.[10]

The new quantum state of superfluid $^3$He indeed corresponds to Cooper pairs but with non-zero angular momenta: orbital quantum number L=1, and spin S=1, both of which are zero in the standard BCS superconductor. Nonetheless, superfluid $^3$He can be understood with a version of this theory described in the book by Vollhardt and Wölfle.[11] The discovery of superfluid $^3$He marks the birth of unconventional superconductivity; more precisely, superfluids that break symmetries of the normal state in addition to gauge symmetry.

The additional quantum degrees of freedom of superfluid $^3$He correspond to broken rotational and time reversal symmetries leading to non-trivial topology. There are two phases in zero magnetic field, A and B. The B-phase is special as it dominates the pressure-temperature phase diagram. Its characteristic is a broken relative spin-orbital rotation symmetry that admits a hierarchy of excited states precisely classified by total angular momentum, J = 0, 1, and 2. The J = 2 states are themselves bosons exactly analogous to the Higgs boson recently observed with the large hadron collider in 2012.[12] These superfluid $^3$He-B excitations were calculated by Maki[13] and Serene[14] although they were first found by Vdovin in unpublished work in 1963.[7,11] It is remarkable that this fundamental broken symmetry of the B-phase is responsible for coherent propagation of shear waves predicted by Moores and Sauls[15] and then confirmed through observation of an acoustic Faraday effect.[16] Transverse sound is unheard of for liquids and is often assumed to be the sole province of rigid solids.

Since the discovery of superfluid $^3$He, many unconventional superconductors have been found. Best known are the cuprates with L=2 and S=0, discovered by Bednorz and Müller in 1986, Nobel prize in 1987, as well as certain heavy fermion compounds.[17] However, only one superconducting compound, UPt$_3$, has multiple superfluid phases analogous to $^3$He with L=3, and S=1 as was predicted.[18] One of its phases breaks time reversal symmetry similar to $^3$He-A.[19] Now the search is on for new superconducting materials, new superfluids, and in the case of superfluid $^3$He, to investigate new phases stabilized by confinement in aerogels (see the figure), in small pores, and narrow slabs.[20]

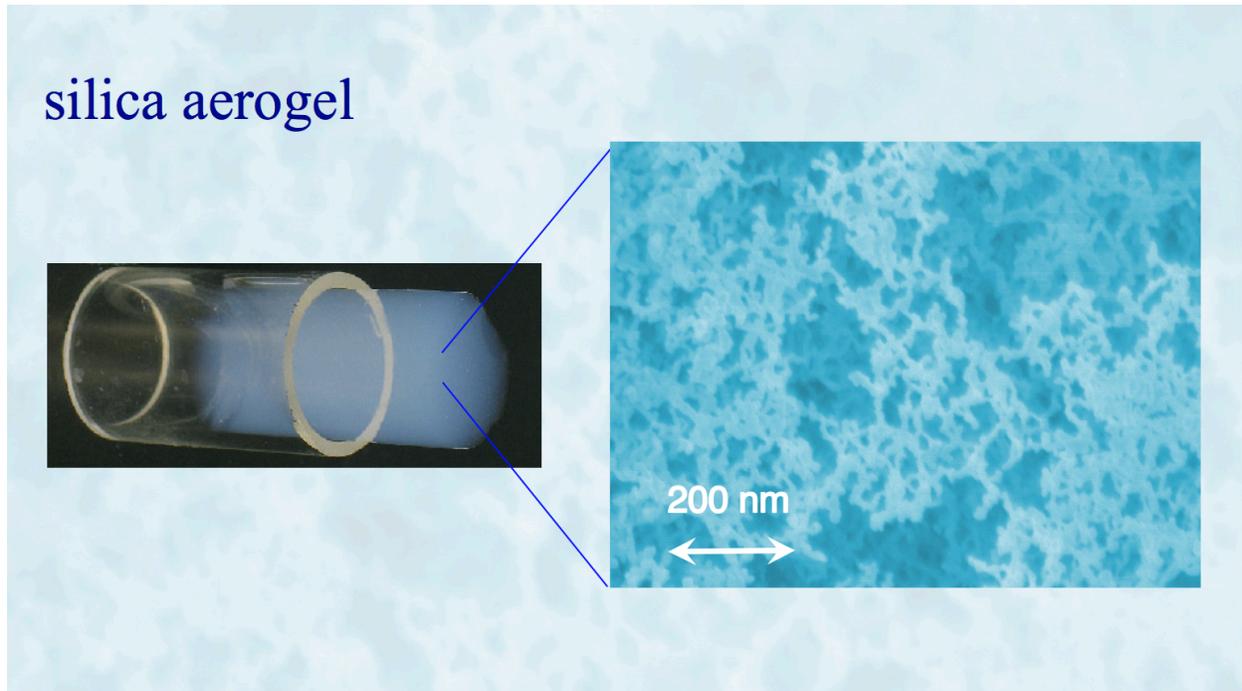

**Figure caption.** High porosity silica aerogel used for confinement of superfluid $^3$He with a scanning electron microscope image at right. The early experiments that revealed superfluidity for the first time by Allen and Misener[1] and Kapitza[2] were conducted on $^4$He in confinement. Recent analogous studies of superfluid $^3$He confined to silica aerogel at 1000 times lower temperatures have revealed new superfluid phases.[20]